\def\year{2022}\relax
\documentclass[letterpaper]{article} % DO NOT CHANGE THIS
\usepackage{aaai22}  % DO NOT CHANGE THIS
\usepackage{times}  % DO NOT CHANGE THIS
\usepackage{helvet}  % DO NOT CHANGE THIS
\usepackage{courier}  % DO NOT CHANGE THIS
\usepackage[hyphens]{url}  % DO NOT CHANGE THIS
\usepackage{graphicx} % DO NOT CHANGE THIS
\usepackage{natbib}  % DO NOT CHANGE THIS AND DO NOT ADD ANY OPTIONS TO IT
\usepackage{caption} % DO NOT CHANGE THIS AND DO NOT ADD ANY OPTIONS TO IT
\DeclareCaptionStyle{ruled}{labelfont=normalfont,labelsep=colon,strut=off} % DO NOT CHANGE THIS
\usepackage{algorithm}
\usepackage{algorithmic}

\usepackage{newfloat}
\usepackage{listings}
\floatstyle{ruled}
\newfloat{listing}{tb}{lst}{}
\floatname{listing}{Listing}

\newcommand\name[1]{\ensuremath{\mathsf{#1}}}

\usepackage{xcolor}
\definecolor{string-color}{rgb}{0.3333, 0.5254, 0.345}
\definecolor{key-color}{rgb}{0.8, 0.47, 0.196}
\title{PantheonRL:\\A MARL Library for Dynamic Training Interactions}
\author{
    Bidipta Sarkar\equalcontrib,
    Aditi Talati\equalcontrib,
    Andy Shih\equalcontrib,
    Dorsa Sadigh
}
\affiliations{
    Department of Computer Science, Stanford University \\
    \{bidiptas, atalati, andyshih, dorsa\}@stanford.edu \\
}

\begin{document}

\maketitle

\begin{abstract}
We present \name{ PantheonRL }, a multiagent reinforcement learning software package for dynamic training interactions such as round-robin, adaptive, and ad-hoc training.
Our package is designed around flexible agent objects that can be easily configured to support different training interactions,
and handles fully general multiagent environments with mixed rewards and \(n\) agents.
Built on top of \name{ StableBaselines3 }, our package works directly with existing powerful deep RL algorithms. Finally, \name{ PantheonRL } comes with an intuitive yet functional web user interface for configuring experiments and launching multiple asynchronous jobs. Our package can be found at \url{https://github.com/Stanford-ILIAD/PantheonRL}.
\end{abstract}

\section{Introduction}

\begin{figure}[t]
    % \begin{minipage}{0.5\textwidth}
    \centering
    \includegraphics[width=.9\columnwidth]{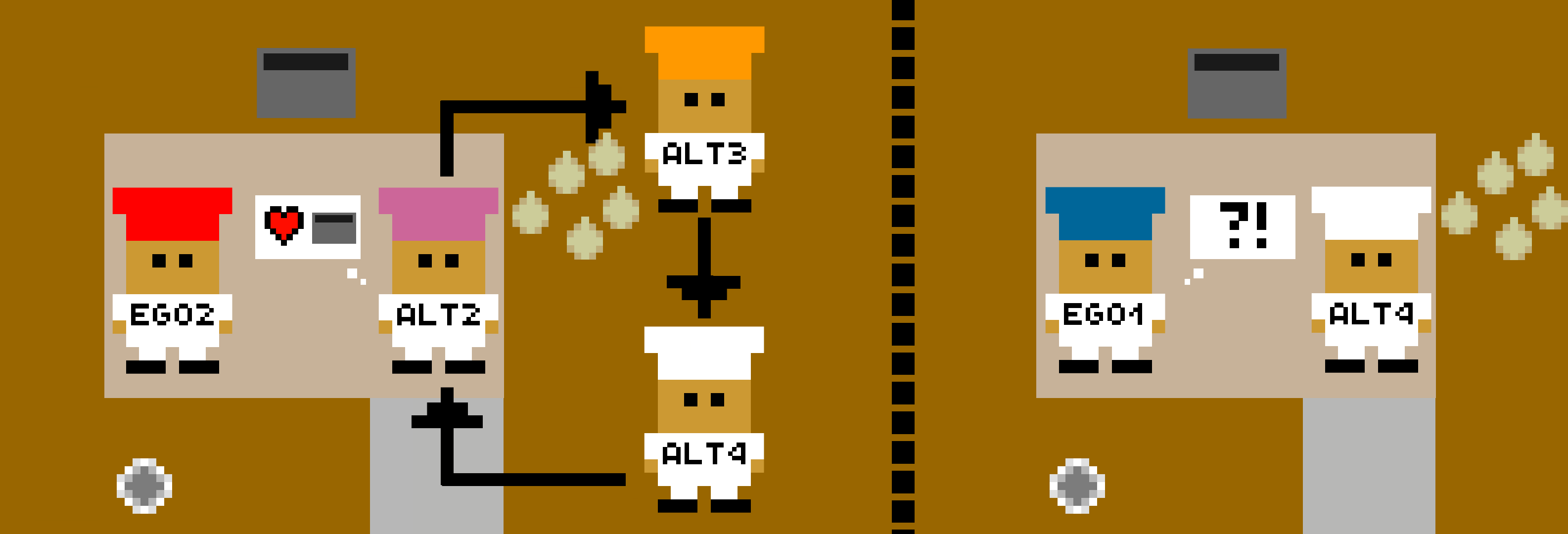}
    \caption{\name{ PantheonRL } is designed to support dynamic training interactions, e.g., round-robin (left) or cross-play (right).}
    \label{fig:cover}
    % \end{minipage}
    % \hfill
    % \begin{minipage}{.45\textwidth}
    %     \centering
    %     \includegraphics[width=0.9\textwidth]{agent_selection_screen_short.png}
    %     \caption{Agent selection screen of the web user interface. A demo of this can be found at \url{https://youtu.be/3-Pf3zh_Hpo}.}
    %     \label{fig:website}
    % \end{minipage}
\end{figure}

Multiagent reinforcement learning (MARL) is becoming increasingly important as more AI systems are being deployed. Many potential applications of MARL involve dynamic interactions between agents, such as agents adapting to each other, ad-hoc coordination, and more (Fig~\ref{fig:cover}). However, experimenting with these dynamic interactions using modern deep RL frameworks can be a difficult process. Existing MARL libraries are largely designed around training a fix set of agents, making them unsuitable for experimenting with more dynamic and adaptive agent interactions. 

We propose \name{ PantheonRL }, an easy-to-use and extensible MARL software package that focuses on dynamic interactions between agents. The goals of our package are:
\begin{enumerate}
    \item to support adaptive MARL, with dynamic training interactions ranging from self-play, round-robin, adaptive (few-shot), and ad-hoc (zero-shot) training,
    \item to build on top of existing powerful deep RL libraries, in particular \name{ StableBaselines3 } (\name{ SB3 })~\cite{stable-baselines3},
    \item to provide a web user interface for launching and monitoring experiments, with support for the different dynamic training interactions described above.
\end{enumerate}

\section{Related Work}

There exists a number of MARL libraries, such as \name{ Mava }~\cite{pretorius2021mava}, \name{ PyMARL }~\cite{samvelyan19smac}, and \name{ PettingZoo }~\cite{terry2020pettingzoo}. The first two offer a selection of centralized multiagent training algorithms (QMIX, VDN, MADDPG), and \name{ PettingZoo } hosts a large collection of multiagent environments. More related to our work is the combination of \name{ PettingZoo } with \name{ RLLib }~\cite{liang2018rllib}, which provides some level of customization of agents. However, to the best of our knowledge, there does not exist a library designed around flexible agent objects for adaptive MARL. In contrast to previous work, our package prioritizes the modularity of agent objects. Each agent object is equipped with its own replay buffer and learning algorithm, allowing users to easily mix and match agents, swap their roles, and finetune individually for adaptive MARL.

\section{PantheonRL Framework}

One of our desiderata is to build upon powerful single agent reinforcement learning (SARL) libraries. Our main design choice is thus: how do we interface with SARL algorithms, which train a policy network with input/output designated by the state/action space of an OpenAI Gym~\cite{openaigym} environment?
Directly running a SARL algorithm on top of a joint multiagent environment (with joint observations and joint actions) is possible, but this produces a monolithic joint policy network that is undesirable for our first desiderata -- to support adaptive MARL with dynamic training interactions. 

Instead, our design is to split the training of each of the \(n\) agents as separate SARL instances, so that we produce \(n\) cleanly distinct policy networks that can be composed  or finetuned for downstream adaptive MARL tasks. We next describe how we designed the environments and agents in a way that supports an intuitive API.

\paragraph{Joint / Projected Environments}

Existing multiagent environments are generally defined as a \emph{joint environment}, with a Gym \texttt{step} function handling \(n\) actions and observations. Each joint environment implicitly defines \(n\) \emph{projected environments} that are Hidden-Parameter MDPs~\cite{Doshi-VelezK16}. The \(i\)-th projected environment handles the \(i\)-th agent's actions and observations, and has hidden-parameters characterized by the policies of the other \(n-1\) agents.

Given a joint environment, we require only specification of the state/action spaces of the \(n\) projected environments. \name{ PantheonRL } then automatically links  with SARL algorithms to produce individual agent policies for each agent.

% All environments in this framework are instances of MultiAgentEnv, which itself is a subclass of OpenAI's gym \name{Env}. Each MultiAgentEnv represents an environment with a fixed number of players, where one of these players is designated as the `ego' agent. 

% Each environment must implement the \name{n\_step} and \name{n\_reset} functions, which are analogous to the single-agent \name{step} and \name{reset} functions. These multi-agent functions specify which players will move in the next turn, their observations, and the rewards of all players at each timestep. There are also two subclasses of MultiAgentEnv that represent common two-player paradigms. SimultaneousEnv is the base class for 2-player environments where both players move at each timestep. TurnBasedEnv is the base class for 2-player environments where players take turns moving in alternate timesteps.

\paragraph{Ego / Partner Agents}

\name{ PantheonRL } differentiates between an ego agent and the other \(n-1\) partner agents. Each agent is equipped with its own replay buffer and learning algorithm. A critical design feature is that each agent's learning algorithm can be chosen from off-the-shelf \name{ SB3 } algorithms, such as PPO, without any modifications.

We distinguish the ego agent role for two reasons. First, when doing round-robin training or adaptation, we often want to fix the ego agent and sample partners from a pool of possible partners. Second, to provide an intuitive API similar to that of \name{ SB3 }, we use the ego agent as the entry-point to the training procedure of all agents. In other words, \name{ego.learn()} will step through the environment, which triggers the learning algorithm of all the partner agents. The triggers are implemented so that all agents reuse the same joint trajectories, to avoid naively collecting the joint trajectories \(n\) times.

We highlight again the importance of \name{ Pantheon }'s compatibility with \name{ SB3 } in giving our framework great flexibility. Each agent can specify its own learning algorithm imported directly from \name{ SB3 }. Designing our API around the ego agent also gives us access to many of the single-agent logging, debugging, and monitoring utilities of \name{ SB3 }.

\section{Web User Interface}

In addition to full-fledged command-line invocations, \name{ PantheonRL } provides an easy-to-use web interface for launching and monitoring experiments. The web interface is valuable for prototyping MARL experiments, especially the dynamic training interactions supported by our package. This minimizes \name{ PantheonRL }'s initial user overhead, which is often non-trivial for other MARL packages. A demo of the user interface can be found at \verb!https://youtu.be/3-Pf3zh_Hpo!.

The website guides the user in selecting the experiment parameters in stages -- first configuring the environment, and then configuring each individual agent. The parameters are presented as dropdown menus, check boxes, buttons, and more, which help reduce a user's mental load during selection. The configurations also allow the user to save/load agent policies and trajectories for later experiments.

Not only is the website visually intuitive, it is also highly functional. One of the main features of the website is its asynchronous design. Built on top of Flask, the website allows a user to launch multiple asynchronous training jobs in the background. After the training jobs have launched, the website can either pull lightweight logging information to display to the user, or spawn a full Tensorboard service in the background to give the user complete monitoring capabilities. Moreover, the website stores a user's session information in a database, and supports login/logout if the user wants to manage multiple sessions.

\section{Dynamic Training Interactions}

Here, we demonstrate the simplicity of our API when training dynamic MARL interactions. The examples are on a \(2\)-player environment, but they extend to \(n\)-player environments just as easily. First, we train an ego agent in a round-robin style by pitting it against two partner agents (Listing 1). Partner 1 is loaded from a previously trained PPO policy, but we wrap it as a \texttt{StaticPolicyAgent} so that its policy does not update anymore. Partner 2 will update its policy using A2C, and the ego agent will update its policy using PPO.

\begin{listing}[tb]%
\caption{Example round-robin python training script}%
\begin{lstlisting}[language=Python, 
    stringstyle = {\color{string-color}},
    keywordstyle = {\color{key-color}},]
env = gym.make('OvercookedMultiEnv-v0')
partner_1 = PPO.load(partner_1_file)
env.add_partner_agent(StaticPolicyAgent(partner_1.policy))
partner_2 = A2C('MlpPolicy', env)
env.add_partner_agent(OnPolicyAgent(partner_2))
ego_a = PPO('MlpPolicy', env, verbose=1)
ego_a.learn(total_timesteps=500000)
\end{lstlisting}
\end{listing}

% \begin{minted}
% [
% framesep=2mm,
% fontsize=\scriptsize,
% ]{python}
% env = gym.make('OvercookedMultiEnv-v0')
% partner_1 = PPO.load(partner_1_file)
% env.add_partner_agent(StaticPolicyAgent(partner_1))
% partner_2 = A2C('MlpPolicy', env)
% env.add_partner_agent(OnPolicyAgent(partner_2))
% ego_a = PPO('MlpPolicy', env, verbose=1)
% ego_a.learn(total_timesteps=500000)
% \end{minted}

Next, we will evaluate partner adaptation by loading a previously trained ego agent, and pairing it with the partner 2 agent we just trained (Listing 2). Since we want the ego agent to adapt to the partner, we set partner 2 to not update its policy. Other training paradigms like ad-hoc pairing and standard MARL can be specified in a similar fashion.

\begin{listing}[tb]%
\caption{Example adaptation script}%
\begin{lstlisting}[language=Python, 
    stringstyle = {\color{string-color}},
    keywordstyle = {\color{key-color}},]
env = gym.make('OvercookedMultiEnv-v0')
env.add_partner_agent(StaticPolicyAgent(partner_2.policy))
ego_b = PPO.load(ego_b_file, env=env)
ego_b.learn(total_timesteps=500000)
\end{lstlisting}
\end{listing}
% \begin{minted}
% [
% framesep=2mm,
% fontsize=\scriptsize,
% ]{python}
% env = gym.make('OvercookedMultiEnv-v0')
% env.add_partner_agent(StaticPolicyAgent(partner_2))
% ego_b = PPO.load(ego_b_file, env=env)
% ego_b.learn(total_timesteps=500000)
% \end{minted}

\name{ PantheonRL } provides a concise API while allowing for great flexibility in customizing training interactions, including the pool of partner agents, the toggling of partner updates, and the training algorithm for each individual agent.

\section{Discussion}

With focus on adaptive MARL and dynamic training interactions, \name{ PantheonRL } is a valuable addition to the MARL software ecosystem. The modularity of the agent policies combined with the inheritance of \name{ StableBaselines3 } capabilities together give users a flexible and powerful library for experimenting with complex multiagent interactions. To top it off, our intuitive yet functional web user interface, equipped with clean visuals and asynchronous job launches, makes \name{ PantheonRL } a suitable library for a wide range of users.

% The general environment framework allows for many different types of multi-player games. The reward signals given by the \name{n\_step} function allow for cooperative, competitive, and mixed-interaction settings.

\section{Acknowledgments}

We would like to acknowledge Office of Naval Research, and Air Force Office of Scientific Research for their support.

\bibliography{ref}

% \begin{minted}
% [
% framesep=2mm,
% fontsize=\scriptsize,
% ]{python}
% env = gym.make('OvercookedMultiEnv-v0')
% alt3 = PPO.load(alt3_file)
% env.add_partner_agent(StaticPolicyAgent(alt3))
% alt4 = A2C('MlpPolicy', env)
% env.add_partner_agent(OnPolicyAgent(alt4))
% ego2 = PPO('MlpPolicy', env, verbose=1)
% ego2.learn(total_timesteps=500000)
% \end{minted}

% Next, we will evaluate partner adaptation by loading a previously trained ego agent, and pairing it with the partner 2 agent we just trained. Since we want the ego agent to adapt to the partner, we set partner 2 to not update its policy.

% \begin{minted}
% [
% framesep=2mm,
% fontsize=\scriptsize,
% ]{python}
% env = gym.make('OvercookedMultiEnv-v0')
% env.add_partner_agent(StaticPolicyAgent(alt4))
% ego1 = PPO.load(ego_b_file, env=env)
% ego1.learn(total_timesteps=500000)
% \end{minted}

\end{document}